\documentclass[useAMS,usenatbib]{mn2e}
\usepackage{graphicx}
\title[Star formation in high redshift radiogalaxies]{Ly$\alpha$ excess in high redshift radio galaxies: a 
signature of star formation.\thanks{Unpublished results are presented
in this paper for the radio galaxy 1338-1941, based on observations carried out at the
European Southern Observatory, Paranal (Chile) for the ESO project 69.B-0078(B).}}

\author[Villar-Mart\'\i n et al.]{M. Villar-Mart\'\i n$^1$, A. Humphrey$^2$,  C. De Breuck$^3$,
R. Fosbury$^4$, L. Binette$^2$,
J. Vernet$^3$\\
$^{1}$Instituto de Astrof\'\i sica de Andaluc\'\i a (CSIC), Aptdo. 3004, 18080 Granada, Spain\\
$^{2}$Instituto de Astronom\'\i a, UNAM, Ap. 70-264, 04510, DF, M\'exico\\
$^{3}$European Southern Observatory, Karl Schwarschild Str. 2, D-85748 Garching bei M\"unchen, Germany\\
$^{4}$ST-ECF, Karl Schwarzschild Str. 2, D-85748 Garching bei M\"unchen, Germany\\
}

\begin{document}

\date{}

\pagerange{\pageref{firstpage}--\pageref{lastpage}} \pubyear{2002}

\maketitle

\label{firstpage}

\begin{abstract}

 About 54\% of radio galaxies at $z\ge$3 and 8\% of radio galaxies
at 2$\la z<$3
 show unusually strong Ly$\alpha$ emission, compared with the
general population of high redshift ($z\ga$2) radio galaxies. These  Ly$\alpha$-excess
objects (LAEs) show  Ly$\alpha$/HeII values
  consistent with or {\it above} standard photoionization model predictions.  

We reject with confidence several scenarios  to explain the
unusual strength of Ly$\alpha$ in these objects: shocks, low nebular metallicities,
high gas densities and absorption/scattering effects. We show that the most successful  explanation is the 
presence of a young stellar population which provides the extra supply of
ionizing photons required to explain the Ly$\alpha$ excess in at least the most extreme LAEs
(probably in all of them).
 This interpretation is strongly
supported by the tentative trend found
by other authors for $z\ge$3 radio galaxies to show lower UV-rest frame polarization levels,
or the dramatic increase on the detection rate at submm wavelengths of $z>$2.5 radio galaxies. 
The enhanced star formation activity in LAEs could be a consequence of a recent
merger which has triggered both the star formation and the AGN/radio activities.

 The measurement of 
unusually high Ly$\alpha$ ratios in the extended gas of some high redshift radio galaxies
suggests that star formation activity occurs in spatial scales of tens of kpc. 

We argue  that, although the fraction of LAEs may be incompletely determined, 
both at 2$\la z<3$ and at $z\ge3$,  the much larger fraction of LAEs found at $z\ge$3 is  a genuine redshift evolution
and not due to selection effects. Therefore,
our results suggest that the
radio galaxy phenomenon is more often associated with a massive starburst at $z>$3 than
at $z<$3.

\end{abstract}


\section{Introduction}

High redshift powerful radio galaxies (HzRG, $z\ga$2) are characterized by an
emission line spectrum which is rich in emission lines of different ionization
species. In the optical (UV rest frame), Ly$\alpha$ is usually the strongest
emission line, followed by CIV$\lambda$1550, HeII$\lambda$1640, CIII]$\lambda$1909 and, less frequently, NV$\lambda$1240 (CIV, HeII, CIII] and NV hereafter).

 De Breuck et al. (2000a, DB00a hereafter) observed that Ly$\alpha$ is unusually strong
in
some HzRG, most of which are at  $z\ga$3.
They  also found that the  Ly$\alpha$ equivalent width  increases at $z\ga3$ and
  the ratios of Ly$\alpha$ to CIV, CIII] and HeII  are roughly
twice the value found for lower redshift objects (2$\la z <3$). 
 The authors interpret these
results as spectroscopic evidence for the youth of radio galaxies at $z\ga$3,
which are still surrounded by large halos of primordial hydrogen from which the galaxy is formed. 

In this paper we present a more thorough study of \cite{breu00a}
 results.
We start by quantifying (\S3)  the  variation of the Ly$\alpha$/HeII and Ly$\alpha$/CIV
 ratios  and Ly$\alpha$ luminosity with redshift for  radio galaxies at 1.8$\le z \le$4.4 and interpret the results
in the light of appropriate photoionization models (described in \S3) and/or scenarios (\S5).
 The results and the cosmological consequences will be 
discussed in \S6. Summary and conclusions will be presented in \S7.

A  $\Omega_{\Lambda} =$ 0.73, $\Omega_{m}$ = 0.27 and $H_{0}$ = 71 km  s$^{-1}$  Mpc$^{-1}$ cosmology is adopted in this paper.

\section{Object sample.}

The object sample has been extracted from the  data set of DB00a and 
De Breuck et al. (2001, DB01 
hereafter), which includes the largest
 compilation of line fluxes for $z\ga$2 radio galaxies available. We will not
consider lower $z$ objects, for which  Ly$\alpha$ is not observable.
Our final sample consists of 13 radio galaxies at $z\ge$3 and 48 radio galaxies at 1.8$\le z <$3. 

Table 1 shows the sample with some relevant
information.

The original data set of De Breuck et al. includes radio galaxies from nine samples designed to find 
HzRG. As the authors explain, the surveys can be divided into two types
(see DB00a for all appropriate references of each survey):
(i) several   large flux density-limited surveys
such as 
the  3CR, MRC, BRL sample, and 6C, 7C and 8C surveys; 
(ii) several ``filtered" surveys, which
 were designed 
to find the highest redshift objects. In the latter case, the radio spectral index
flux is most often used (ultra steep spectrum sources), sometimes in
combination with an angular size upper limit. Alternatively, the filter
consists of a flux density interval centered around the peak in the source
counts around 1 Jy.

The total sample contains
176 radio galaxies with $z$ in the range 0.05 to 5.2. It contains
a similar number of sources at $z<2$ and  at $z>$2.
For the present study, we have eliminated $\sim$110 objects for the
following reasons:

\begin{itemize}

\item 4 broad  emission line objects (FWHM$\ga$4000 km s$^{-1}$ and/or
additional quasar-like properties).
These are
  2313+4053 ($z=$2.99, \cite{breu01}), 0744+464 ($z=$2.93, McCarthy \citeyear{mc91}),
 2025-218  ($z=$2.63, Villar-Mart\'\i n et al. \citeyear{vm99}, Larkin et al. \citeyear{lar00}) and 0349-211  ($z=$2.33, McCarthy et al. \citeyear{mc91b}).
For these objects, the UV lines have a contribution
from the
 broad line region, where the physics  is complex and not interesting for
the current study.

\item 2 objects with broad absorption line systems:
1908+722 ($z=$3.53) and 1115+5016 ($z=$2.54) (\cite{breu01}).

\item 91 objects with no  Ly$\alpha$ measurement. Except
for 3 objects  for which the reason is not known,
 in all other cases Ly$\alpha$ is outside 
the observed spectral range. 

\item 23 objects  with no measurement of CIV {\it and} HeII.
The reasons are varied: 
for 7 objects, the CIV and HeII lines were outside the useful 
spectral range, or only Ly$\alpha$ narrow band imaging fluxes are available.
For 6 objects, the lines are reported to be
detected, but no fluxes are provided, except for Ly$\alpha$. 
For 7 objects, the lines where within the observed spectral range,
but not detected. In 2 cases HeII was outside the observed spectral range
and CIV was inside, but not detected. For 1 object, the reason for
the non-availability of the CIV and HeII fluxes  is unknown. 

Using \cite{breu01} original spectra, 
 we have been able to measure useful upper limits (in some cases the
spectra are too noisy or  severely affected by sky residuals)
for the CIV and/or HeII lines of several of these objects.
The lower limits for the Ly$\alpha$ ratios are shown in Table 1.

\end{itemize}

There are 10 objects  with no measurement of CIV (2)
{\it or}  HeII (8). In all cases, except one, both lines where
within the observed spectral range, but not detected. 
Upper limits
could be measured for two objects using the original \cite{breu01}
spectra. The lower limits for the Ly$\alpha$ ratios 
are given in Table 1.

The following objects were part of De Breuck et al. sample,
but the line fluxes of one or more of the lines
of interest to us were not provided. For this reason, we present measurements based on our own 
long slit Keck LRISp spectroscopic data or results published by other authors.

\begin{itemize}

\item 4C40.36  at $z=$2.27 (nuclear aperture, Humphrey \citeyear{hum05}, 
Humphrey et al. 2006, in prep; Vernet et al.~\citeyear{ver01})
)

\item 4C48.48  at $z=$2.34 (nuclear aperture, Humphrey \citeyear{hum05}, 
Humphrey et al. 2006, in prep; Vernet et al.~\citeyear{ver01})

\item  0731+438 at $z=$2.43
(nuclear aperture,  Humphrey \citeyear{hum05}, 
Humphrey et al. 2006, in prep; Vernet et al.~\citeyear{ver01})

\item 0902+34 at $z=$3.40 (Mart\'\i n-Mirones et al. \citeyear{miron95})

\item 1243+036 at $z=$3.56 (nuclear aperture,  Humphrey \citeyear{hum05}, 
Humphrey et al. 2006, in prep; Vernet et al.~\citeyear{ver01})

\item  4C41.17 at $z=$3.79 (Dey et al. \citeyear{dey97})

\item  1338-1941  at $z=$4.11 (De Breuck et al. in prep.
ESO project 69.B-0078(B).)

\end{itemize}

More accurate line measurements than De Breuck et al. original data were
used for:

\begin{itemize}

\item 2334+1545 at $z=$2.48.  HeII is not detected in this object.
Because of the unusually  narrow CIV 
line ($<$500 km s$^{-1}$) measured by \cite{breu01}
compared with Ly$\alpha$ (1900 km s$^{-1}$)
and CIII] (1600 km s$^{-1}$) lines, we checked this spectrum again.
We cannot confirm that the line is detected. We conclude that Ly$\alpha$/CIV$\ga$5
and Ly$\alpha$/HeII$\ga$5.

\item  2104-242  at $z=$2.49 (based on  VLT
  FORS long slit spectroscopy, Overzier et al. \citeyear{over01})

\item 0140+3253 at $z$=4.41. According to \cite{breu01},
CIV is detected  and Ly$\alpha$/CIV=43$\pm$10.
Because of the strong sky residuals in the CIV region, we have checked the spectrum again.
We cannot confirm the detection of the line. We have measured
upper limits for both CIV and HeII. We obtain   Ly$\alpha$/CIV$\ga$30
and Ly$\alpha$/HeII$\ga$10.

\item Upper limits for CIV and/or HeII were measured for
0231+3600 ($z$=3.08), 0040+3857 ($z$=2.61), 2355-002 ($z$=2.59) and
2254+185 ($z$=2.15) using the original spectra of \cite{breu01}. 

\end{itemize}

\begin{table*}
\begin{tabular}{lllllllll} \hline
  Object & $z$ &  Ly$\alpha$/HeII &  Ly$\alpha$/CIV &   P(\%) & $S_{850\mu m}$ (mJy)  \\ \hline
 {\bf  0140+3253} & 4.41      &  $\ga$10  & {\bf $\ga$30} & -- & 3.3$\pm$1.5 \\
{\bf 1338-1941} &  4.11   & {\bf 18$\pm$4}  &  11$\pm$2 & 5.0 &  7$\pm$1  \\ 
4C60.07 & 3.79   & --  & 3.7      & -- &  17$\pm$1 \\ 
{\bf 4C41.17}  & 3.79  &    {\bf  27$\pm$2}  & 11.1$\pm$0.7  & $<$2.4 & 12.1$\pm$0.9  \\ 
{\bf 1911+6342}  & 3.59     & {\bf 18$\pm$2}  & 18$\pm$2 & -- & $<$11.9 \\ 
{\bf 2141+192} & 3.59    & {\bf 17.8}  &  10.6  & --  &  $<$5   \\ 
{\bf 1243+036}  & 3.57     & {\bf 34$\pm$3}  & 14$\pm$1 & 11$\pm$4 & $<$6.5  \\ 
0121+1320  & 3.52   &  7.1$\pm$2  & 8$\pm$2  & 7$\pm$2  & 7.5$\pm$1.0    \\ 
{\bf 0205+2242} &  3.51 & {\bf 26$\pm$4}  & 18$\pm$2 &  -- & $<$5.2    \\ 
0902+34 &  3.40  &    9.1  & 6.3 & 1.5  & --  \\ 
1123+314  & 3.22   &     2.6$\pm$0.4 & 1.8$\pm$0.3 & --  & 5$\pm$1   \\ 
1112-2948  & 3.09   &     2.4$\pm$0.3 &  4.1$\pm$0.7  & -- &  5.9$\pm$1.6 \\
0231+3600  & 3.08   &     $\ga$5 &  $\ga$5  & -- & 6$\pm$2 \\ \hline
0747+365 & 2.99    &    3.5$\pm$0.7  & 4.6$\pm$0.9 & --  & 5$\pm$1   \\ 
0943-242 &  2.92  &    7.4  &  5.2 & 6.6$\pm$0.9  & --   \\ 
4C24.28	 &  2.88  &    --  & 4.3 & -- &  2.4$\pm$1.2    \\ 
0857+036 &  2.81  &    3.7  & 2.6  & --  & --  \\ 
0417-181 &  2.77  &    6.0  & 2.5 &  --& --   \\ 
1019+0534 &  2.77  &    1.0   & 0.8  & -- &   $<$3.8  \\ 
1338+3532 &  2.77     & 5.9$\pm$0.8  & 14$\pm$3 & --  & -- \\ 
{\bf 0920-071} & 2.76     & {\bf 15$\pm$2}  & 12$\pm$2  & -- & --  \\   
1545-234 &  2.76  &    3.7  & 6.1 & -- &  --  \\ 
2202+128 & 2.71   &    3.2  & 4.1 & -- & --   \\ 
1357+007  & 2.67   &    --  & 5.4  & --  &  --  \\ 
0040+3857 & 2.61  &    $\ga$5   & 4$\pm$2 & -- &  -- \\ 
2355-002	 & 2.59      & $\ga$3 & 4.0$\pm$1 &  -- &  --    \\ 
0529-549 &  2.58  &    12.3  & 18.5 & --  &   --  \\  
0828+193 &  2.57  &     7.0 & 7.0 & 10$\pm$1 &  --  \\ 
{\bf 1755-6916} & 2.55     &  {\bf 23$\pm$4}  & 15$\pm$2 & --  & --   \\ 
1436+157 & 2.54   &   7.0  & 2.5 & -- & --   \\ 
1558-003 &  2.53  &     8.8  & 5.5  & --  & --   \\ 
{\bf 0303+3733 } & 2.51     & {\bf 21$\pm$4}  & 7$\pm$1 & -- & --     \\ 
1650+0955 & 2.51   &    8$\pm$1  & 7$\pm$1 & -- &  --  \\ 
2104-242 & 2.49   &    10$\pm$1  & 13$\pm$2 & --  & --    \\
2334+1545 & 2.48 & $\ga$5 & $\ga$5 & -- & --  \\
4C23.56	 &  2.48  &    5.3  & 3.8 & 15$\pm$2 & $<$5.5    \\ 
2308+0336 &  2.46     & 7.5  & 4.7 & --  & $<$3.2   \\ 
{\bf 0731+438} &  2.43     & {\bf 14$\pm$1}  & 8.6$\pm$0.7 & $<$2.4 & --  \\ 
1033-1339 & 2.43   &    12$\pm$2  &  4.3$\pm$0.7 &--  &  --   \\ 
0748+134 &  2.42  &    4.2  & 3.5 & -- &  --   \\ 
1410-001 &  2.36  &    11.1  & 8.3 & 12$\pm$3 & --    \\ 
1707+105 &  2.35  &   4.8 &   --  & -- & --    \\ 
0128-264 &  2.35     &   --  & 7.2  & --  & --   \\ 
1428-1502	 &  2.35    & 9$\pm$2  & 9$\pm$2 & &  \\
4C48.48  &  2.34  &    11$\pm$1  & 8.2$\pm$0.8 &   8.4$\pm$1.5  & 5$\pm$1  \\ 
0211-122 &  2.34  &    1.0  & 1.8 & 19$\pm$1 &     \\  
1251+1104 & 2.32   &    7.7  & 7.7 & -- &  --   \\ 
4C40.36	 &  2.27  &    11.6$\pm$0.8  & 7.3$\pm$0.7 & 7$\pm$1 &  $<$4.1   \\ 
1113-178 & 2.24   &    9.1  & 3.8 & -- & -- \\ 
0200+015  &  2.23  &    5.4  & 4.1 & -- & -- \\ 
1138-262	 &  2.16     & 10.7  & 17.4 & -- & 13$\pm$3  \\ 
0355-037 &  2.15  &    3.0  & 4.1 & -- &  --  \\ 
2254+185 &  2.15  &    10$\pm$5  & $\ga$4.0  & $\la$2.6 & --   \\ 
2036+0256 & 2.13   &  10$\pm$3    & 11$\pm$5  & -- & --   \\ 
1102-1651 & 2.11   &    2.1$\pm$0.5  & 2.7$\pm$0.7 & -- & --   \\ 
1802+645 & 2.11   &    --  & 8.9 & --  & --  \\ 
1740+664 &  2.10  &    --  & 13.7 & --  &  --  \\ 
0448+091 & 2.04   &    8.7  & 10.1 & -- &  --  \\  
1805+633 &  1.84  &     -- & 6.8 & --  &  -- \\ 
3C256 &  1.82  &    9.9  & 10.4 & 12$\pm$2  &  --   \\ 
3C294 & 1.79   &    10  & 10 & -- &  $<$2.3  \\ 
 \hline
\end{tabular}
\caption{Sample of radio galaxies used in this study 
  in order of decreasing redshift and with some relevant information. LAEs are highlighted
in bold. $z\ge$3 and $z<3$ radio galaxies are separated by a horizontal line.
See \S2 for references of the optical
data. P(\%) values are from Vernet et al. \citeyear{ver01}, Dey et al. \citeyear{dey97}, De Breuck
et al. (in prep.) and Jannuzi et al. \citeyear{jan95}. 
Submm data are from Archibald et al. \citeyear{arch01} and Reuland et al. \citeyear{reu04}. No available
measurements are indicated with '--'. Errors for P(\%) and $S_{850}$ are provided when available.
Upper limits for $S_{850\mu m}$ are 3$\sigma$ values.} 
\end{table*}

\section{Models}

Photoionization models were computed using the multipurpose  code Mappings Ic 
(Binette, Dopita \& Tuohy \citeyear{bin85}, Ferruit et al. \citeyear{fer97}). The following assumptions
will be applied unless stated otherwise:  the ionizing continuum is a
power law (PL) of index $\alpha$=-1.5 ($F_{\nu}\sim \nu^{\alpha}$) with a cut off energy
of 5$\times$10$^4$ eV. The gas is characterized by a  density  $n=$100 cm$^{-3}$ and solar abundances.
The clouds are radiation bounded and the geometry is plane
parallel. The density behaviour within the clouds is isobaric. Dust reddening effects
have been ignored in the modeling. These are
the standard models which, in general, reproduce most of  the UV and optical line ratios of HzRG
rather successfully (e.g. Robinson et al. \citeyear{rob87}, Villar-Mart\'\i n et al. \citeyear{vm97}, Humphrey \citeyear{hum05}).

All the stellar SEDs (Spectral Energy Distributions) used in this paper were  built
 using Starburst99 (Leitherer et al. \citeyear{lei99}),  a web based software and data package designed to model spectrophotometric and related properties of star-forming galaxies.  
We have assumed an initial mass function (IMF) with spectral indexes 1.3 and 2.3 for
the mass intervals 0.1 to 0.5 M$_{\odot}$ and 0.5 to 120 M$_{\odot}$
respectively. Stellar abundances are always considered to be solar.
 Both continuous and instantaneous star forming histories were considered.

\section{Variation of   L\lowercase{y}$\alpha$/H\lowercase{e}II, L\lowercase{y}$\alpha$/CIV and L\lowercase{y}$\alpha$ luminosity with  redshift}

In this section we investigate how the
Ly$\alpha$/HeII and Ly$\alpha$/CIV ratios and the Ly$\alpha$ luminosity
change  with redshift.
Both ratios  have been measured in many radio galaxies at
 $z\ga$2, since the lines are observable in the optical window.

Three diagnostic diagrams involving the  Ly$\alpha$, HeII and CIV lines
are shown in Fig.1. The solid line represents standard photoionization models
(see \S3).
 The ionization parameter
$U$\footnote{$U=\frac{Q}{4\pi r^2nc}$, where $Q$ is the photon ionizing
luminosity, $r$ is the distance between the cloud and the ionizing source, $n$ is the
particle gas density and $c$ is the speed of light.}  varies within the range [0.0046,1].
 
The Ly$\alpha$/HeII ratio predicted by the models is in the range $\sim$15-20. 
Due to its strong sensitivity to $U$, the predicted Ly$\alpha$/CIV values
cover a much larger range. The minimum predicted value is $\sim$9. Because Ly$\alpha$/HeII
shows little dependence on $U$ compared with Ly$\alpha$/CIV, we will focus most of our discussion
on the first ratio, with reference to other  ratios when useful.
  
All objects in Table 1 with measurements available for  the three
line ratios (including lower and upper limits) are plotted in Fig. 1. 
 We have marked with coloured symbols  those radio galaxies with  Ly$\alpha$/HeII values
 consistent  with or above the
standard case B  value (0140+3253 at $z=$4.41 is the only exception, see below). Given that errors are not available for
many radio galaxies in the sample, we consider an object to be consistent with
the model predictions when   Ly$\alpha$/HeII is   at least 90\% of the minimum values predicted
by the models, i.e. Ly$\alpha$/HeII$\ge$14.   Because of the
very high values compared with typical values of most HzRG, we will call these radio galaxies Ly$\alpha$
excess objects (LAEs hereafter).
Black symbols correspond to objects with Ly$\alpha$/HeII ratios
 below the case B recombination value.

The diagrams show  a very clear segregation of the most distant radio galaxies
in the sense that most $z\ge$3 objects   have the highest Ly$\alpha$/HeII 
values.  The Ly$\alpha$/CIV ratios
of these objects are also among the largest values.

Of the 10 sources at $\ge$3 for which Lya/HeII has been measured, 6 have
 Ly$\alpha$/HeII$\ge$14
and are therefore LAEs. 4 are not LAEs. 
 It is not possible to
classify the remaining three $z\ge$3 radio galaxies
 due  to the unavailability of the Ly$\alpha$/HeII value.

One more $z\ge$3 radio galaxy 
can be added to the group of LAEs:  0140+3253 ($z=$4.41), which 
shows Ly$\alpha$/HeII$\ga$10 and  Ly$\alpha$/CIV$\ga$30.
The large Ly$\alpha$/CIV ratio does not necessarily imply that the object
is a LAE, since  this ratio can be reproduced by  very low
or  high $U$ values (see Fig.~1). The real value of Ly$\alpha$/HeII  would be necessary.
However, the fact that this object presents  the largest Ly$\alpha$/CIV 
value of
the whole sample  suggests that
this is also a LAE in the sense that the Ly$\alpha$ 
emission is unusually enhanced. This is also suggested by its high
Ly$\alpha$ luminosity (1.35$\times$10$^{44}$ erg s$^{-1}$), which is more  characteristic
of LAEs than non LAEs (see below).  Therefore, we classify 0140+3253
as a LAE.

We conclude that at least 7 out of 13 (54\%) 
 $z\ge$3 radio galaxies
are LAEs.  On the other hand, only 4 out of
48 (8\%) 2$\la z<$3 radio galaxies can be classified as such
(see Table 1)\footnote{Considering a power law of index $\alpha$=-1.0 (Villar-Mart\'\i n, Tadhunter \& Clark, 1997) would not change our
conclusions. Ly$\alpha$/HeII would be in the range $\sim$10-14, implying that still the vast majority of 2$\la z <$3 objects
lie below the model predictions, while most $z\ge$3 radio galaxies are LAEs.}.

We show in Table 2 the median  and  average (1st number in brackets) 
values of the Ly$\alpha$ ratios for the
two redshift samples together with other interesting information
(upper and lower limits have been excluded in the calculations). 
 Ly$\alpha$/HeII and Ly$\alpha$/CIV  
are $\sim$2.2 and $\sim$1.5 times larger in  $z\ge$3 radio galaxies, which also
tend to have noticeably larger Ly$\alpha$  ($\sim$3 times) luminosities  compared with the
low $z$ sample. The CIV/HeII ratio is similar in both groups. The bottom line shows
the probabilities $P$ that the two groups are drawn from different parent populations
according to the Kolmogorov-Smirnov test. 
These values confirm that  $z\ge$3 radio galaxies  show larger Ly$\alpha$ ratios
and higher  Ly$\alpha$ luminosities. Interestingly, there is no distinction in  radio size
between the two samples.

\begin{table*}
\begin{tabular}{cccccc} 
\hline
 $z$ range   & $\frac{Ly\alpha}{HeII}$ &  $\frac{Ly\alpha}{CIV}$ & $\frac{CIV}{HeII}$ & L(Ly$\alpha$) & LAS \\ 
        &           &             &             & 10$^{44}$ erg s$^{-1}$ \\  \hline
 2$\la z <$ 3    &     8.00 (8.32,[4.75])  &  6.90 (7.24,[4.21])    & 1.20 (1.29,[0.65])  & 0.44 (0.74,[0.79]) & 52 (66.2,[60.6])  \\ 
$z\ge$3   &  17.50 (16.01,[10.90])     &   10.60 (10.07,[5.41]) & 1.44 (1.51,[0.59]) & 1.35 (1.34,[1.03]) & 43.0 (78.0,[74.7])   \\  
$P$         & 99\%   & 96.5\%  &  71\%  & 93\%  & 1\%   \\  \hline
\end{tabular}
\caption{The median, average (first number in brackets) and standard deviation
(2nd number in brackets)  of the main UV
line ratios and the Ly$\alpha$ luminosities in the 2$\la z <$3 and $z\ge$3
samples are shown. The Ly$\alpha$/HeII and Ly$\alpha$/CIV ratios are
$\sim$2.2 and 1.5 times higher in the  $z\ge$3 sample. On the other
hand, Ly$\alpha$ is $\sim$3 times more luminous in this sample. No distinction is
apparent on radio size between the two samples.  The bottom line 
shows the  probability
that the two samples are drawn from different parent populations according to the Kolmogorov-Smirnov test.
(Radio largest angular size values (LAS, in Tables 2 and 3) taken from 
DB01, DB00a and De Breuck et al. 2000b)}
\end{table*}

\begin{figure*}
\includegraphics{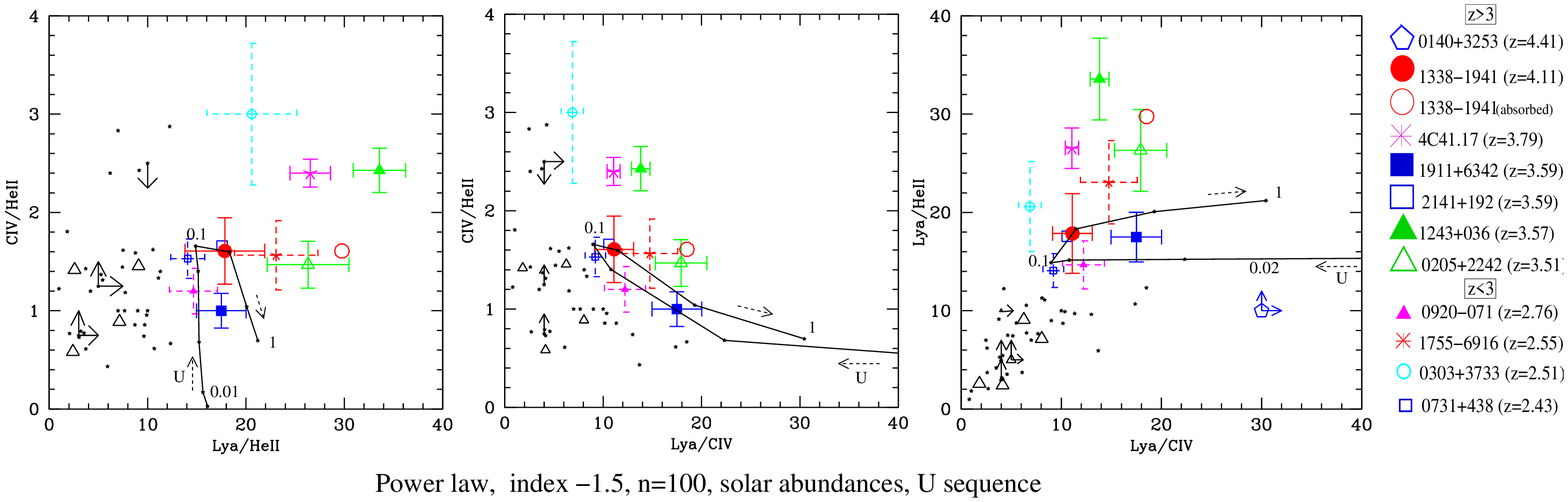}
\vspace{3in}
\caption{Diagnostic diagrams involving the   Ly$\alpha$, HeII and   CIV
emission lines (see electronic
manuscript for colour version of the figures). The solid black line
represents the standard $U$ sequence of photoionization models  (see text),
with the $U$ value indicated for some points along the sequence.
The dashed arrows show the sense of increasing $U$. The 
Ly$\alpha$/HeII values predicted by the models are in the range $\sim$15-20.
Objects with  Ly$\alpha$/HeII  ratios consistent with or above  such values 
(named LAEs throughout this work) 
are indicated in colour and their names shown in the adjacent labels.
 Error bars for these objects are shown when available.
Radio galaxies at $z<$3 are shown with dashed errorbars and/or small sized  symbols 
and radio galaxies
at $z\ge$3 are shown with solid errorbars and/or large sized symbols.  Solid arrows indicate lower
and upper limits.
The Ly$\alpha$ ratios have not been
corrected for absorption effects, except for 1338-192  for which both the
absorbed (red solid circle) and corrected (red hollow circle) values are shown.
  The $z\ge$3 sample
tend to have  larger Ly$\alpha$/HeII and Ly$\alpha$/CIV ratios. Ly$\alpha$/HeII
is above the model predictions for several objects. }
\end{figure*}

We have repeated the calculations from
Table 2, this time separating the sample into LAE and non-LAE sources,
rather than into $z<$3 and $z\ge$3 sources. The results are shown in Table 3.
 LAEs tend to be at higher redshift with a median  $z$ value
 $z_{med}$=3.57, while $z_{med}$=2.48 for objects with no Ly$\alpha$ excess.
They have Ly$\alpha$ luminosities $\sim$3.4 times higher and Ly$\alpha$/HeII and Ly$\alpha$/CIV ratios
2.7 and 2.1 times higher respectively compared with non-LAEs. Interestingly,
also the CIV/HeII ratio is 1.6 times higher. There is also a clear
difference in radio size.  The bottom line shows
the $P$ values resulting from  the Kolmogorov-Smirnov test.

\begin{table*}
\begin{tabular}{ccccccc} \hline
Object class & $z$    & $\frac{Ly\alpha}{HeII}$ &  $\frac{Ly\alpha}{CIV}$ & $\frac{CIV}{HeII}$ & L(Ly$\alpha$)& LAS  \\ 
        &           &         &    &             & 10$^{44}$ erg s$^{-1}$ & kpc  \\  \hline
non-LAEs & 2.48 (2.53,[0.42])   & 7.25 (6.94,[3.30]) &    5.45 (6.70,[4.05])      & 1.00 (1.20,[0.60]) &  0.43 (0.77,[0.84]) & 85.2 (53.5,[74.7]) \\
LAEs & 3.57 (3.35,[0.68])    &   19.50 (21.38,[6.20])   &  11.50 (12.52,[3.69])    & 1.63 (1.80,[0.62])  & 1.45 (1.55,[1.05]) & 18.5 (38.8,[42.6])\\ 
$P$ & 98\% & -- & 99.9\% & 99.5\% & 94\% & 90\% \\ \hline
 \end{tabular}
\caption{As Table 2, but the objects have been classified as LAEs (radio galaxies with Ly$\alpha$ excess) and non-LAEs. The LAEs tend to be at higher redshift. Ly$\alpha$ is $\sim$3.4 times more luminous in these objects and the Ly$\alpha$/HeII and Ly$\alpha$/CIV ratios are 2.7 and 2.1 times 
higher than in  non-LAEs. The radio size is $\sim$4.6 times smaller for LAEs.  The bottom lines shows the $P$ values
resulting from the  Kolmogorov-Smirnov test.}
\end{table*}

Interestingly, there are 4 LAEs (out of 11, i.e. 36\%)  which, even after taking  into account the errors, show a Ly$\alpha$/HeII
well {\it above} the maximum value (20) predicted by the models.
These objects are: 1338-1941 (Ly$\alpha$/HeII$\sim$30 once corrected 
for absorption, Wilman et al. \citeyear{wil04}), 4C41.17, 1243+036, 0205+2242, all at $z>$3.
No or very weak absorption has been found for the LAEs  4C41.17, 1243+036 (van Ojik et al. 1997) and 1755-6916 
(Wilman et al. 2004).
This information is not available for the rest of the LAEs.

\subsection{Measurements of spatially extended Ly$\alpha$ ratios}

The results above are based on   spatially integrated spectra,
centered on the high surface brightness regions 
(although not necessarily covering the whole extension of the emission line regions).
There are two $z\ge$3 objects in the literature for which the spatial variation of the Ly$\alpha$ ratios
has been investigated in the direction of the radio structures (both of them
are in the sample above): 1243+036 (green solid triangle in the figures) at $z=$3.36  
(Humphrey \citeyear{hum05}, Humphrey et al. 2006, in prep.) and 2141+192 (blue open square  in the figures) at $z=$3.59 
(Maxfield et al. \citeyear{max02}). 
For 1243+036 the Ly$\alpha$/HeII ratio has values in the range
  26$\pm$5 up to $\ga$40 in the outer parts ($\ga$7'' from the nuclear region) of the object.
Ly$\alpha$/HeII=33$\pm$1
in the nuclear
region.  On the other hand,
Ly$\alpha$/CIV presents a minimum in the nuclear region (14$\pm$1) and a maximum of $\ga$40 (this
is a lower limit) in the
outer parts of the object ($\ga$7'').  The spatially integrated values are $\sim$34$\pm$1 and 14$\pm$1 respectively.

The second object is 2141+192 at $z=$3.59 (Maxfield et al. \citeyear{max02}). The integrated spectrum
gives $\sim$18 and 11 for Ly$\alpha$/HeII and Ly$\alpha$/CIV respectively.  Fig.~7 in \cite{max02}
 shows a region
at $\sim$1-1.5$\arcsec$ from the continuum centroid, where Ly$\alpha$/HeII$\ga$40 
and Ly$\alpha$/CIV$\sim$15-20.  At $\sim$4-4.5$\arcsec$,  Ly$\alpha$/HeII$\ga$30.

\cite{hum05} (Humphrey et al. 2006, in prep.) studied the spatial variation of the Ly$\alpha$  ratios  for a sample of 10 radio galaxies (all included in our study) in the redshift range 2$<z<$3.  Five of these
objects (1558-003, 4C40.36, B3 0731+438, 4C48.48 and 4C-00.54)  show  Ly$\alpha$/HeII in the range 14-30 at some spatial
positions in the extended gas. 
These large Ly$\alpha$/HeII values are measured in all cases outside the nuclear region.
Among these five objects, only B3 0731+438 is
classified as a LAE from its spatially integrated spectrum
(blue hollow small square in the figures).

\section{Scenarios}

In this section, we will investigate $a)$ why Ly$\alpha$ is unusually
strong in LAEs relative to other emission lines
and the continuum, compared with the majority of HzRG;
$b)$ why LAEs show Ly$\alpha$/HeII values which are often well above the standard
model
predictions (see \S4).

We are looking for a mechanism that can enhance the  Ly$\alpha$ emission over other
emission lines and the continuum. We have considered two  
possibilities: 
different  properties of the ionized nebulae
(heavy element
abundances, gas density, absorption and scattering properties) and
 alternative ionization mechanisms (shocks, stellar
photoionization vs.
AGN photoionization).

The diagrams in Fig.~1 show a large scatter of the measured Ly$\alpha$ ratios, 
which
cannot be explained by the standard models. Most radio galaxies show too faint 
Ly$\alpha$ relative to CIV and HeII. The most natural explanation is
 that the Ly$\alpha$ photons are  absorbed by neutral gas 
and dust. Therefore, the most realistic way to model the line ratios
of the {\it whole sample} should take absorption effects into account.
This information is available only for a few objects (Wilman et al. 2004).
Van Ojik et al. (1997) found clear signs of absorption in 11 out of 18 radio galaxies
at $z\ga$2, but they do not provide the  information necessary to correct
the Ly$\alpha$ fluxes in the absorbed objects.

Our expectation is that LAEs should show stronger absorption than non LAEs.
         \cite{breu00a}  found that  Ly$\alpha$ absorption
 occurs most frequently in sources with small radio sources  
(see also van Ojik et al. \citeyear{vo97}) and at higher redshifts. 
Since most LAEs have higher $z$ and smaller radio sizes than non LAEs (see Table 3), 
it is likely that they are also more absorbed. 
In this case, they
are likely to be  more overluminous in Ly$\alpha$ than assumed in this work.

On the other hand,  the goal of this study is to explain the {\it Ly$\alpha$ excess}
in the LAEs. For this and the above reason, absorption will be ignored  in  
our models, but only discussed qualitatively when necessary.  It is,
therefore, not to be expected that the models presented in this section should
reproduce the line ratios of the whole sample, but only those of the LAEs.

\subsection{Lower metallicities.}

We investigate next whether lower nebular metallicities in
  LAEs  can explain the observational results.

A decrease of the heavy element abundances will enhance the
nebular electron temperature, which will favour
the collisional excitation of Ly$\alpha$ (and heavy element lines 
 such as CIV$\lambda$1550), while this process will not affect
HeII$\lambda$1640 or the continuum significantly.

We have computed a set of photoionization models to investigate the impact of the metallicity variation on the Ly$\alpha$ ratios.
  The density has been fixed  to $n$=100
cm$^{-3}$. 
We show in Fig. 2 the same diagnostic diagrams as before, with two model sequences for  $U=$0.05 (dashed black line) and $U=$0.015
(red solid line) respectively. A range of $U$ values is considered, 
since the ionization level of the gas is known to vary from object to
object (i.e., Robinson et al. \citeyear{rob87}, Villar-Mart\'\i n et al. \citeyear{vm97}, Humphrey \citeyear{hum05}).  The metallicity range covered by the two sequences is  0.05 to 2 
$Z_{\odot}$. $U$ values outside the quoted range would produce $a$) similar results 
($U$=0.1 would
produce similar results to $U$=0.05, although implying even lower metallicities
for the objects) or $b$) discrepant line ratios. For example, very low ionization models ($U\la$0.005)
 produce too faint CIV emission so that 
CIV/HeII$<$0.4. 
On the contrary, very high ionization
models ($U\ga$0.3) which can explain Ly$\alpha$/HeII$\ga$25 predict too strong
CIV, with 
CIV/HeII$\ga$4.

\begin{figure*}
\includegraphics{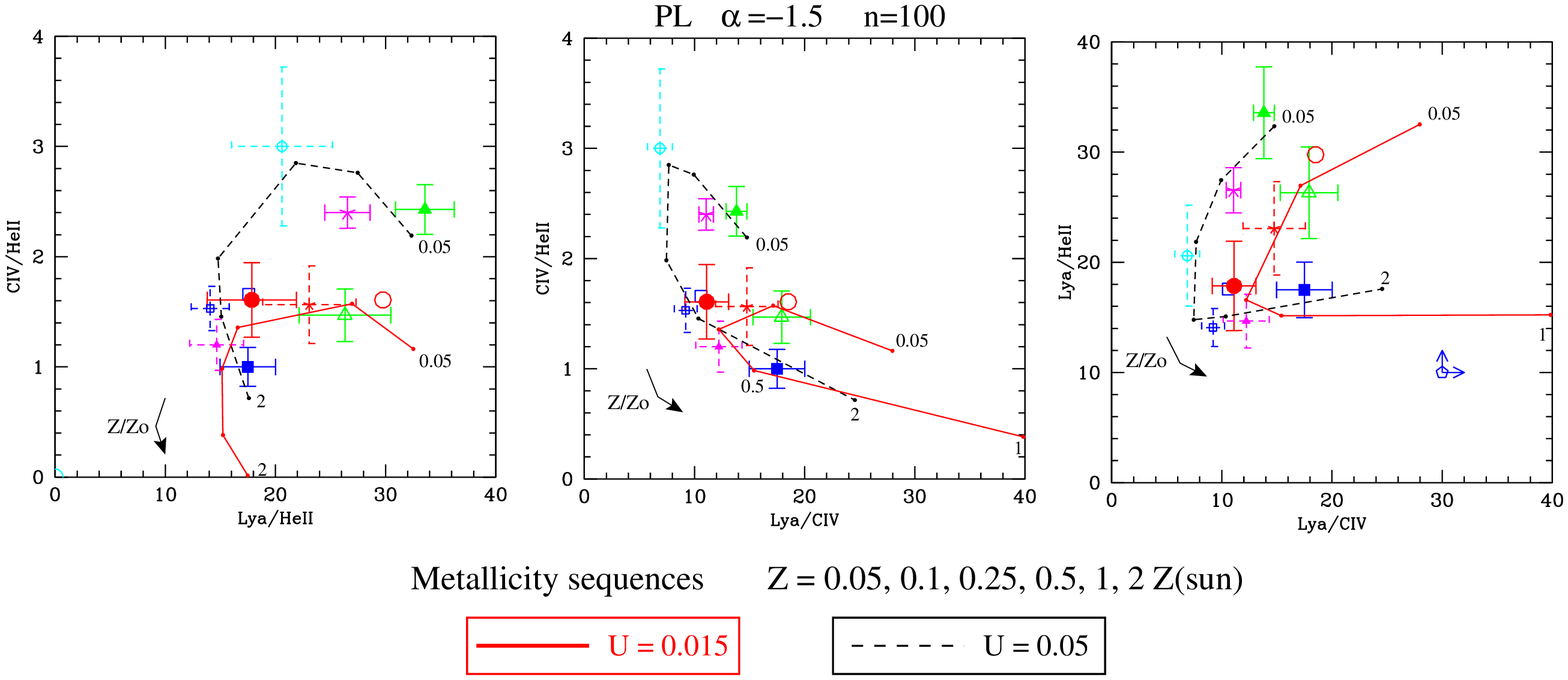}
\vspace{3in}
\caption{Metallicity effects. Same diagnostic diagrams as in Fig.1. The black
and red lines represent sequences
of photoionization models for two different $U$ values. 
 The metallicity varies in the range 0.05 to 2 $Z_{\odot}$
in both cases. The black arrows indicate the sense of increasing metallicity.
Metallicity effects are very successful at reproducing the position of  LAEs  in
these three diagnostic diagrams.  Notice that at least  7 out of 13 objects require abundances $\la$25\% $Z_{\odot}$. However, the models fail to reproduce
the strength of the NV emission in at least the most extreme LAEs (see text).}
\end{figure*}

As the metallicity decreases towards well below solar values,
the  UV line ratios (in particular Ly$\alpha$/HeII) change dramatically. For $Z\ga$50\% $Z_{\odot}$
this ratio is rather constant, since the nebula is too cold to produce any
noticeable collisional excitation of Ly$\alpha$. As the metallicity
decreases
further, the high electron temperature makes this mechanism very
efficient and the Ly$\alpha$/HeII ratio increases by a factor of 2 in
the
metallicity range 5-50 \% $Z_{\odot}$. The models also succeed to
reproduce the high CIV/HeII ratios for some LAEs.

A u-turn in the model sequences is observed for the ratios involving
CIV due to the relatively less efficient excitation of the line (compared
with Ly$\alpha$ and HeII) as the ions
become more and more scarce as the metallicity decreases.
Interestingly, metallicities as
low as 5\% still produce strong metal (CIV) lines relative to HeII, with 
CIV/HeII$>$1 if the
gas
is highly ionized ($U\sim$0.05-0.1).

5  LAEs  are consistent with solar
abundances, choosing the appropriate $U$ value and taking errors into account
(lower metallicities are not discarded, though, due to the $U$-$Z$ degeneracy in this area
of the diagrams). These objects are: 1911+6342 , 2141+192 and 1338-1941 at $z>$3
and  0920-071, 0371+438 at $z<$3. Once corrected for Ly$\alpha$ absorption, one of
them (1338-1941,
red circles) shifts towards the low metallicity models (10\%  $Z_{\odot}$).
 I.e. if
absorption was taken into account, it is possible that more objects
would
require well below solar abundances.

Including 1338-1941,  6  LAEs    have $Z\la$25\% $Z_{\odot}$  according to  their UV line ratios.
 Interestingly, very low abundances $Z\la$10\% $Z_{\odot}$ are inferred for 4 of these objects: 
 1338-1941, 1243+036,  4C41.17, 1338-1941 (once corrected for absorption) and 0205+2242, all at $z>$3.

According to these results,  LAEs are, on average,
characterized
by lower abundances than radio galaxies with no Ly$\alpha$ excess. Abundances of less than 25\% $Z_{\odot}$ 
are implied in $\sim$55\% of  LAEs. Since Ly$\alpha$ absorption has been taken into account
for only one object, the metallicities could be even lower.

The NV$\lambda$1240 line provides a strong test for the validity of
 the low metallicity scenario. According to the models, this line should be very
faint in the most metal poor objects. Its flux has been measured in
three
of them: 1243+036, 4C41.17
and 1338-1992. We show in Table 4 the predictions of the  NV/HeII
 and NV/CIV ratios for the  models that best reproduce the  Ly$\alpha$ ratios of
 these three objects, together  with the measured values.
It is clear that in all three cases the NV emission is observed to be much (2 to 4 times) stronger  than
expected from the model predictions.
Nitrogen overabundance has been frequently found in different types of objects. This 
is usually attributed to 
secondary production of nitrogen. However, this mechanism cannot work at the very low metallicities
implied by the models (Vila-Costas \& Edmunds \citeyear{vila93}).  Therefore, these LAEs are unlikely
to have  low metallicities.

We therefore conclude that low metallicities do not provide a satisfactory explanation for
at least the most extreme LAEs.

\vspace{0.5cm}

\begin{table*}
\begin{tabular}{lllllllllll} \hline
 Object & $\frac{Ly\alpha}{HeII}_{obs}$ & $\frac{Ly\alpha}{CIV}_{obs}$   & $\frac{NV}{HeII}_{obs}$ &   $\frac{NV}{CIV}_{obs}$ & ($U,Z$)$_{mod}$ & $\frac{Ly\alpha}{HeII}_{mod}$ & $\frac{Ly\alpha}{CIV}_{mod}$   & $\frac{NV}{HeII}_{mod}$ &   $\frac{NV}{CIV}_{mod}$ \\   \hline
1243+036$^a$ & 34$\pm$3 & 14$\pm$1 &  0.9$\pm$0.1         & 0.37$\pm$0.06      &  (0.05,5\%$Z_{\odot}$) & 32.3 & 14.8 & 0.26    &  0.12   \\ 
4C41.17$^b$ &   27$\pm$2 & 11.1$\pm$0.7 &  0.71$\pm$0.08       &   0.29$\pm$0.03     &  (0.05,10\%$Z_{\odot}$) & 27.5 & 10.0 & 0.37        &  0.13        \\ 
1338-1992$^c$ &  18$\pm$4 & 11$\pm$2 &  0.29$\pm$0.05           &  0.18$\pm$0.02       &  (0.015,10\%$Z_{\odot}$) & 16.6 & 12.2 & 0.08    &    0.06             \\ \hline
\end{tabular}
\caption{Comparison between the Ly$\alpha$/HeII, Ly$\alpha$/CIV,
NV/HeII and NV/CIV measured ($obs$) and predicted ($mod$) values for three of the most extreme LAEs. Measurements have been taken from: $^a$Humphrey \citeyear{hum05} (Humphrey et al. 2006, in prep.); $^b$Dey et al. \citeyear{dey97}; 
$^c$De Breuck et al. (in prep.). The best model characteristic parameters
 ($U,Z$)$_{mod}$ for each object are shown in column 6. NV is observed to be much stronger
than the models predict.}
\end{table*}

\subsection{Density enhancement.}

An efficient way to enhance  Ly$\alpha$ over other emission lines
and the continuum is increasing the
gas density. At low densities the cooling of the nebula is due mostly to some forbidden
lines (e.g. [OIII]$\lambda$4959,5007, [OII]$\lambda$3727, etc), via  radiative
de-excitation of their upper energy levels in the original ion.   
As the density increases, the critical density will be reached for some
of these transitions, so that de-excitation becomes collisional, rather than
radiative and radiative cooling  becomes much less efficient. The electron
temperature rises as a consequence and  
  Ly$\alpha$ will have
 an increasing contribution of collisional excitation in the partially
ionized zone, which does not affect the HeII$\lambda$1640
 recombination line.

In addition, when densities approach values $\sim$10$^5$ cm$^{-3}$,
 the fraction of energy emitted as Ly$\alpha$ photons versus two-photon continuum
emission due to the
decay  of the 2 $^2$S level of H 
 tends toward unity, rather than 0.67 at the low
 density
limit (Binette et al. \citeyear{bin93}). 

 The dramatic increase of Ly$\alpha$/HeII with density
can be clearly seen in Fig. 3, where the  red-solid and black-dashed lines 
are sequences
of photoionization models for two different $U$ values 
(solid red line: $U$=0.05; black dashed  line: $U$=0.7), with $n$ varying along
each sequence in the range  10$^{1-7}$   cm$^{-3}$ and 10$^{1-6}$   cm$^{-3}$
respectively. CIV collisional excitation is also more  efficient
so that CIV/HeII increases dramatically with density, while  Ly$\alpha$/CIV
decreases.

 Fig. 3 shows that density 
effects (coupled with a variation in $U$) can explain the location of all 
LAEs in the diagnostic diagrams (intermediate $U$ values
would lie between the two sequences). It is important to note that high
densities not only help to reproduce large Ly$\alpha$/HeII ratios, but 
also the trend for LAEs to show larger CIV/HeII, well above standard model
predictions in some cases (see Fig.1).
There are however arguments that make this explanation fail.

The implied densities for some LAEs
 (e.g.  1243+036 and 4C41.17) 
 are $\sim$10$^5$ cm$^{-3}$, i.e. very high compared with the
densities measured or inferred for the extended
emission line regions of radio galaxies at all redshifts  ($\la$several  to few hundred cm$^{-3}$). On the other hand, such high densities  exist in the nuclear narrow line region (e.g. De Robertis \& Osterbrock \citeyear{dero84}).

Is it possible that the emission line spectrum of  LAEs is
dominated by the high density nuclear gas?
This could  be  the case for  one of them,
1338-1941 ($z=$4.11, Ly$\alpha$/HeII$\sim$30, once corrected for absorption), for which 
the two dimensional Ly$\alpha$ spectrum appears very compact (De Breuck
et al. \citeyear{breu99}), or objects such that the aperture used
to extract the spectra studied in this paper was optimized to isolate the
 nuclear emission.
However, very high densities ($\ga$10$^5$ cm$^{-3}$) at large nuclear distances are implied by 
the enhanced Ly$\alpha$/HeII  ratios measured in the extended gas of several
objects (see \S4.1).

An alternative possibility is that the gas that emits most of the
line flux is distributed in very dense 
clumps, which can be at large distances 
 from the nuclear region.  As an
example,  \cite{dey97} find that 30\% of the Ly$\alpha$ flux in 4C41.17 comes from 
extranuclear clumps
in the scale of a few hundred pc. However, the implied photon 
ionizing luminosities would be unrealistically high: $Q\sim$10$^{59-61}$  s$^{-1}$ for
 $r\sim$10 kpc,  
compared with   $\sim$several$\times$10$^{57}$ erg s$^{-1}$ for  $z\ga$2 quasars (e.g. Heckman et al. \citeyear{heck91a}).

We  conclude that, in general, high densities do not provide a natural explanation
for the Ly$\alpha$ excess in  LAEs.

\begin{figure*}
\includegraphics{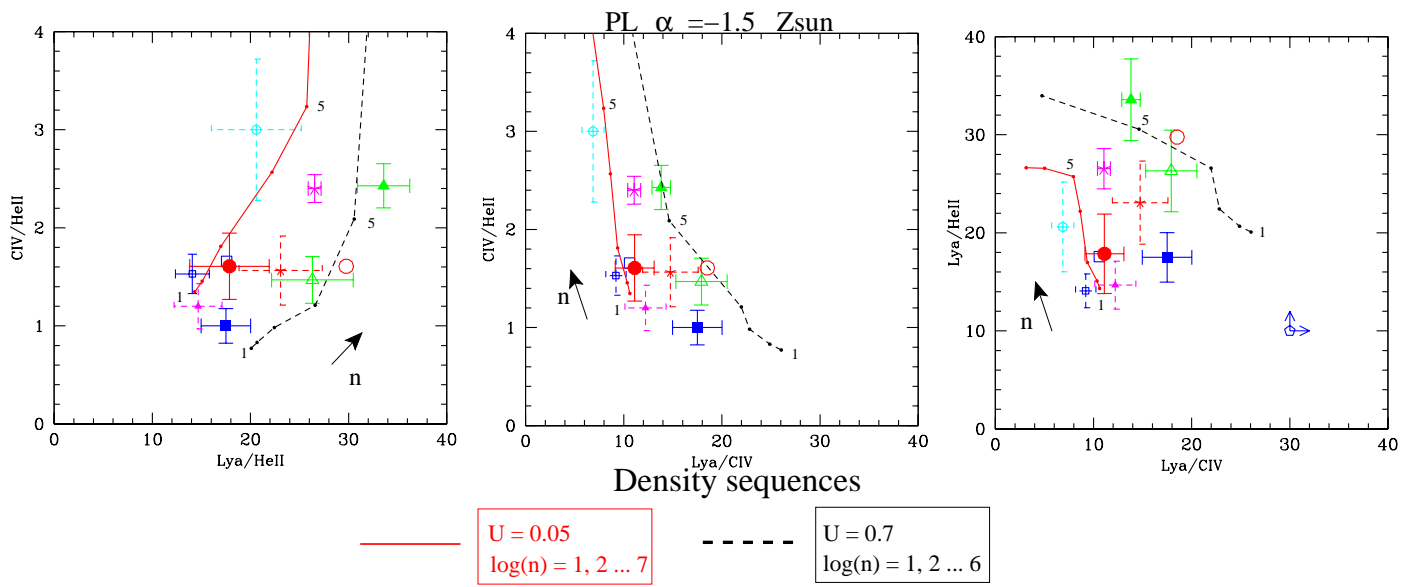}
\vspace{3in}
\caption{Density effects. Same diagnostic diagrams as in Fig.1 with two
model sequences for different $U$ values.  
Solid line: $U=$0.05;  dashed line: $U$=0.7. The density $n$  varies along each sequence
between $n$=10 and 10$^7$ cm$^{-3}$ or 10$^6$ cm$^{-3}$ respectively. Density 
effects (coupled with a variation in $U$) can explain the location of all 
LAE in the diagnostic diagrams. However, unrealistically high
 densities  ($n\sim$10$^5$ cm$^{-3}$) are required to reproduce the line ratios of
the highest Ly$\alpha$ excess objects (see text).}
\end{figure*}

\subsection{Absorption and resonant scattering}

The simplest explanation for the larger Ly$\alpha$ ratios in  LAEs
compared with objects with no Ly$\alpha$ excess is that the line is less absorbed. This would also explain why
Ly$\alpha$ is also more luminous and has larger equivalent widths. It does
not explain, however, why Ly$\alpha$/HeII is above the standard model predictions
in some cases.

 A way to enhance Ly$\alpha$ over other emission lines is by means of resonant scattering,
with  an absorbing 
medium  distributed in very dense and compact neutral dusty clumps
(Hansen \& Oh \citeyear{ho06}). In such scenario,
the Ly$\alpha$ photons can be scattered by the gaseous surface of the neutral clumps without reaching the dust. All  other emission lines and continuum can penetrate the dusty cores,
where they will suffer absorption. If the medium is quite clumpy,
the Ly$\alpha$ photons will be able to escape unabsorbed.
This can produce a noticeable increase of the line equivalent width and
the ratio relative to other lines. \cite{vm96} proposed a similar explanation  for  the existence of pure Ly$\alpha$ emitting regions in some radio galaxies. It has also been suggested by Hansen \& Oh  (\citeyear{ho06}) to explain the intriguing large EW Ly$\alpha$ values measured in some Ly break galaxies. \cite{ver01} also found that in order to get an efficient  grey scattering that
explains the continuum polarization properties of HzRG,
 a high contrast clumpy medium is needed.

In this scenario,  the ambient medium around LAEs consists of dense neutral dusty
clumps while objects with no Ly$\alpha$ excess are surrounded by a more diffuse dusty medium
where Ly$\alpha$ is very efficiently quenched.
However, as we explain above (\S5) LAEs are likely to be more strongly absorbed than non-LAEs.

\subsection{Shocks}

So far we have investigated scenarios such that the same 
mechanism ionizes the gas in LAEs and non-LAEs, but the nebular properties
are different.  An alternative possibility is that the nebulae have similar
properties, but the ionizing
mechanism is different in both types of objects.

Shocks could enhance the Ly$\alpha$ emission via collisional excitation of the line,
thanks to the  heating effect and/or density enhancement. Two of the  LAEs with the highest
Ly$\alpha$ excess (e.g. 1243+036 and
4C41.17)
show, indeed, clear signs of jet gas interactions (van Ojik et al. \citeyear{vo96},
Bicknell et al. \citeyear{bick00}).  LAEs in general tend to
have small radio sizes and broad emission lines, also suggestive of jet-gas interactions (e.g. Humphrey et al. \citeyear{hum06a}, Best et al. \citeyear{best01} ).

The median values of the Ly$\alpha$ luminosity for  LAEs and
non-LAEs are 1.45$\times$10$^{44}$ and 0.43$\times$10$^{44}$ erg s$^{-1}$
 respectively. Therefore, ignoring absorption,  LAEs
produce  an excess of Ly$\alpha$ luminosity given by 
$L(Ly\alpha)_{exc}\sim$(1.45-0.43)$\times$10$^{44}$ erg s$^{-1}$=10$^{44}$ erg s$^{-1}$.
There is, of course,
a dispersion   around $L(Ly\alpha)_{exc}$, however, in order 
to simplify the argumentation that follows, we
will consider the above value  as the characteristic Ly$\alpha$ excess
for  LAEs over typical radio galaxies with no Ly$\alpha$ excess.

If shocks are responsible for the  Ly$\alpha$ excess,
it must be so that the other emission lines are not noticeably
 enhanced. Shock models with velocities in the range 200-1000 km s$^{-1}$, solar
abundances and a gas precursor density of $n=$1 cm$^{-3}$
produce Ly$\alpha$/HeII in the range $\sim$25-100 and  Ly$\alpha$/CIV in the range $\sim$10-100 for certain magnetic parameter values within the range $B$=10$^{-4}$-10, which
depend on the shock velocity (Mark Allen, private communication). For shock velocities
larger than 400 km  s$^{-1}$, Ly$\alpha$/H$\beta$ is in the range 30-80. I.e.,
the shock cooling gas can be characterized by an emission line spectrum
dominated by the Ly$\alpha$ line.
 
 Shock models with  Ly$\alpha$/HeII in the range 25-35 (as measured for the
most extreme LAEs, i.e. those with the highest Ly$\alpha$ excess) produce Ly$\alpha$/H$\beta\sim$35. Therefore, the expected 
H$\beta$ luminosity  from the shocked gas  $L(H\beta)_{shock}$ is
 2.9$\times$10$^{42}$
erg s$^{-1}$.  $L(H\beta)_{shock}$   is related to the shock
velocity in the clouds $V_w$ and the mass flow rate through the shock
$\dot M$ by the following equation (adapted from equation 4.4 of Dopita \& Sutherland \citeyear{ds96}):

$$ L(H\beta)_{shock} = 2.8 \times 10^{37} ~[\frac{V_w}{100~km~s^{-1}}]^{1.41}
\frac{\dot M}{M_{\odot}~yr^{-1}} ~erg ~ s^{-1}$$

Assuming  $V_w$=500 km~s$^{-1}$,  this implies 
$\dot M \sim$10$^4$ M$_{\odot}$ yr$^{-1}$. If the hot spots advance speed
is in the range 3$\times$10$^{3-4}$km s$^{-1}$ (e.g. Scheuer \citeyear{sch95}), 
the radio structures will need $\sim$1.7$\times$10$^{6-7}$ yr to cross
the
$\sim$50 kpc radius nebulae. The total amount of material consumed
in the shock will be $\sim$1.7$\times$10$^{10-11}$ M$_{\odot}$, which is similar
or  larger
than the mass of the giant nebulae (e.g. McCarthy  \citeyear{mc93}, 
Villar-Mart\'\i n et al. \citeyear{vm03}) and is unrealistic.

If we assume that  $L(H\beta)_{shock}$ shows the same dependence on $V_w$ for
values $>$500
km s$^{-1}$ (although we do not know whether this is the case), extreme shock velocities as high as $V_w\ga$5000 km s$^{-1}$
are required to obtain reasonable  mass flow values
$\la$several$\times$10$^8$ M$_{\odot}$.

The continuum emitted by the shocked gas can also photoionize the precursor gas ahead of the shock
(e.g. Dopita \& Sutherland \citeyear{ds96}, Bicknell et al. \citeyear{bick00}). 
However, for photoionization of the precursor  to enhance Ly$\alpha$ efficiently
over the other emission lines, very high densities and/or low metallicities
would be required, as for standard photoionization (see above). As we
discussed in \S5.1 and \S5.2, both scenarios have problems.

Finally, we cannot explain either why Ly$\alpha$ enhancement is not observed in the low $z$ sample, where the extreme effects of shocks
are evident in many objects (e.g. Villar-Mart\'\i n et
al. \citeyear{vm03}, Humphrey et al. \citeyear{hum06a}).

We conclude that shocks do not provide a natural explanation for the
Ly$\alpha$ enhancement in the  LAEs.

\subsection{Stellar photoionization}

An efficient way to enhance  Ly$\alpha$  over other emission lines
  is by means of stellar photoionization of the nebula, thanks to the
soft ionizing continuum.

 Assuming that 67\% of the ionization processes
end up in a Ly$\alpha$ photon (Binette et al. \citeyear{bin93}), $L(Ly\alpha)_{exc}\sim$10$^{44}$ erg s$^{-1}$
corresponds to a total ionizing luminosity of $Q_{ion}^{abs}\sim$10$^{55}$   s$^{-1}$
{\it absorbed} by the nebula.

A continuous burst of star formation
with a  star forming rate (SFR) of  200 $M_{\odot}$ yr$^{-1}$ produces $Q_{ion}^{mod}$=(0.9, 1.7, 2.7)$\times$10$^{55}$
s$^{-1}$ for ages of 1, 2 and 5 Myr respectively, consistent with $Q_{ion}^{abs}$   taking into account that
a fraction of the photons can escape the nebula and Ly$\alpha$ might have suffered some absorption.\footnote{The
escape fraction has very uncertain values. It can vary
between $\sim$0.03 in some nearby starburst galaxies (Leitherer et al. 1995)
and $>$0.5 in Lyman break galaxies (Steidel, Pettini \& Adelberger \citeyear{ste01}).} Values up to a few  thousands M$_{\odot}$ yr$^{-1}$ in HzRG have been inferred from submm studies
(e.g. Archibald et al. \citeyear{arch01}).

Let us compare the expected continuum flux at the redshifted 1550 \AA\ with the observed values
for two of the most extreme LAEs: 4C41.17 ($z=$3.79) and 1243+036 ($z$=3.57) for which
 $f_{1550}$=1.3$\times$10$^{-18}$  and 3.4$\times$10$^{-19}$  erg s$^{-1}$ cm$^{-2}$ \AA\ s$^{-1}$
respectively in the observer's frame
(Dey et al. \citeyear{dey97}); Humphrey \citeyear{hum05}, Humphrey et al.  2006,  in prep.).

The expected monochromatic fluxes (nebular plus stellar, with the nebular
component being $\la$20\% of the total continuum luminosity)  of the bursts discussed above at the same wavelength
 are:
(0.4, 0.8 and 1.6)$\times$10$^{-18}$ erg s$^{-1}$ cm$^{-2}$ \AA$^{-1}$ for
1, 2 and 5 Myr respectively at $z\sim$3.7, which
are consistent with the measured values.

 For an instantaneous (rather than continuous) burst of star formation with the same IMF,
 a burst of age $\la$2   Myr must have a mass 3.8$\times$10$^8$ M$_{\odot}$ to reproduce
${Q^{abs}}_{ion}$
(older ages require larger masses).
The expected monochromatic flux at 1550 \AA\ (at $z\sim$3.7)
is $f_{1550}$=5.1$\times$10$^{-19}$ erg s$^{-1}$ cm$^{-2}$ \AA\ $^{-1}$ in 
reasonable good agreement with
the continuum level of 1243+036. A mass $\sim$10$^9$ M$_{\odot}$ would be required to
explain the continuum level of 4C41.17 (for this object, no contribution from scattered continuum is
expected, Dey et al. \citeyear{dey97}).

 It remains to be explained
the trend for LAEs to show  higher  CIV/HeII compared with non-LAEs
(median values 1.7 and 1.0 respectively, see Table 3). This ratio is  well above
the standard power law model predictions for several objects (see Fig.1).

A  nebula photoionized by stars can be characterized by very large CIV/HeII$>$1000
 (Villar-Mart\'\i n et al. 2004). On the other hand, provided that
the gas is highly ionized,  CIV can have a non-negligible flux compared with Ly$\alpha$,
with Ly$\alpha$/CIV$\ga$10 . Therefore, the UV emission line spectrum of such
a nebula will be dominated by Ly$\alpha$ and CIV, while the contribution to HeII (and NV) will be negligible.
This is for instance the case of the Lynx arc, a gravitationally lensed  star forming galaxy at $z=$3.36
(Fosbury et al. \citeyear{fos03}, Villar-Mart\'\i n et al. \citeyear{vm04}). 

We show in Fig. 4 the UV line ratio predictions of models  which consider an increasing contribution of stellar
photoionized gas relative to the AGN ionized gas (i.e. relative to the standard power law model predictions
shown in Fig. 1). The $\omega$ parameter represents the ratio between the Ly$\alpha$ flux
emitted by the stellar ionized gas and that emitted by the AGN  photoionized gas. Three values have been considered
$\omega$=0 (i.e. pure AGN photoionization, black dashed line), 0.7 (red solid line) and 1.5 (green dotted line).
For illustrative purposes, the stellar model predictions used  here are those presented in Table 6 of Villar-Mart\'\i n
et al. 2004, 
corresponding to a single stellar age of 5 Myr and $U=$0.2 for the stellar photoionized gas. 
A Salpeter IMF in the mass range 2-120 M$_{\odot}$ and an instantaneous star forming history
were assumed.
 Ly$\alpha$/H$\beta\sim$26 for this model.
Fig. 4 shows that a varying contribution of stellar photoionized  gas to the emission line spectrum of LAEs is successful at reproducing the position of these objects  in
these three diagnostic diagrams.

\begin{figure*}
\includegraphics{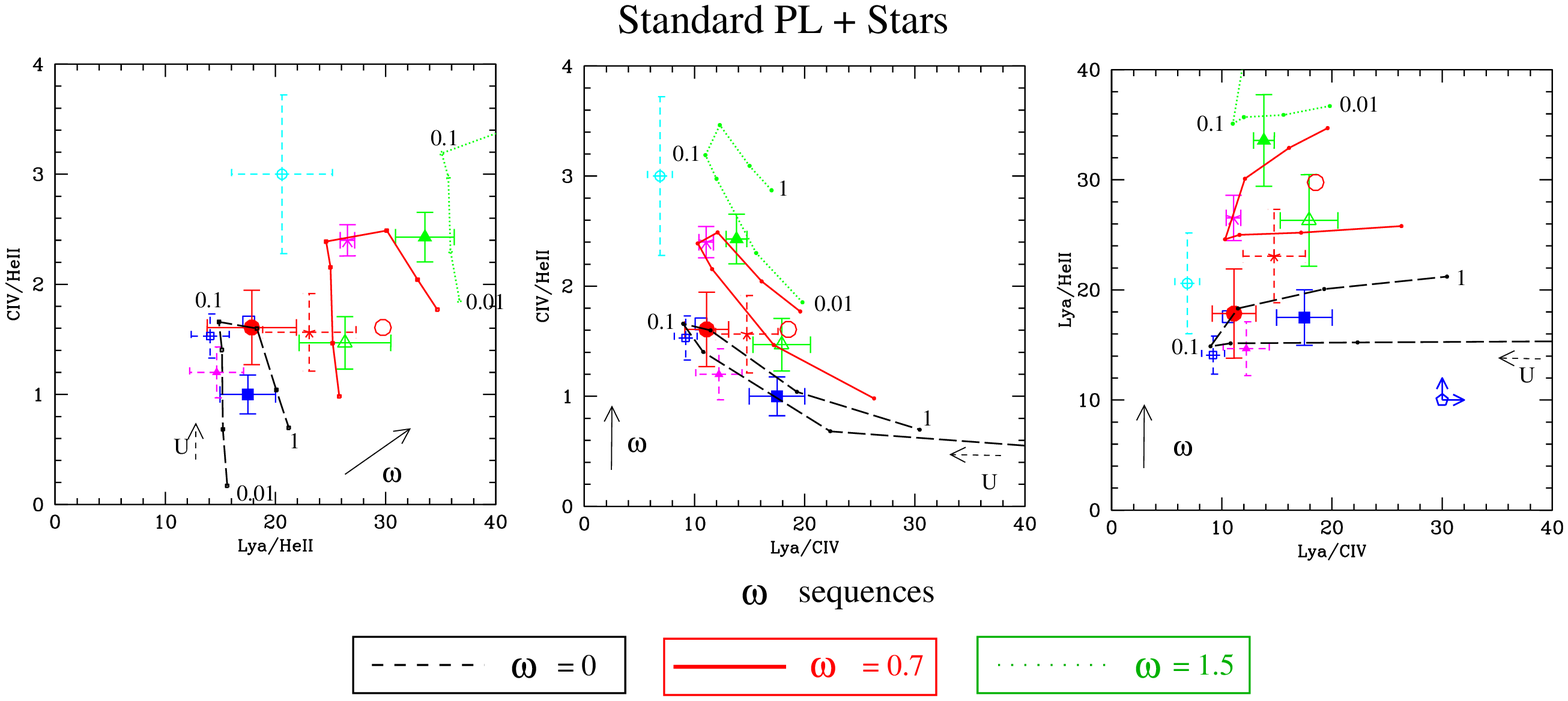}
\vspace{3.2in}
\caption{Effects on the UV line ratios due to the additional contribution of stellar photoionized gas.
$\omega = \frac{Ly\alpha_{*}}{Ly\alpha_{PL}}$ represents the ratio between the Ly$\alpha$ flux
emitted by the stellar ionized gas and that emitted by AGN  ionized gas.
Same diagnostic diagrams as in Fig.1. The black dashed
line represents the standard PL $U$ sequence (see Fig.1 ), i.e., the spectrum due to 
pure AGN photoionization ($\omega$=0).
The coloured lines represent models  with an increasing contribution of emission
  lines from gas ionized by stars to the emission
line spectrum, $\omega$=0.7 and 1.5 for the red solid and green dotted lines respectively (see text for more
details).
$U$ values  for the PL sequence are shown for some models. 
 The black solid arrows indicate the sense of increasing stellar contribution.
A varying contribution of stellar ionized gas to the emission line spectrum of  LAEs is very successful at reproducing the position of these objects  in
these three diagnostic diagrams. }
\end{figure*}

Stellar photoionization  is, therefore, a promising explanation 
for the Ly$\alpha$ excess in LAEs.
 A continuous star forming history with a star
forming rate of $\sim$200 M$_{\odot}$ yr$^{-1}$ is very successful  at explaining the excess of Ly$\alpha$ luminosity
and the observed level of continuum
in at least the two most extreme LAEs.  Since other objects show similar
continuum levels  and less extreme Ly$\alpha$ excesses, it is likely that this assertion is valid for all LAEs.

45\% of LAEs require $\omega>0$: 1338-1941, 4C41.17, 1243+036 and 0205+2242 ($z>$3) and
0303+3733 ($z<$3).  The remaining LAEs are consistent with $\omega$=0 within
the errors (i.e. star formation is not required).  However, since no information on the fraction of absorbed Ly$\alpha$ flux is available,
except for 1755-6916 (which is not absorbed, Wilman et al. 2004),
$\omega$=0 is a lower limit.

\section{Discussion}

There is additional evidence that supports stellar photoinization as the mechanism
responsible for the Ly$\alpha$ excess. 
Fig.~5 shows Ly$\alpha$/HeII (in log) vs. $P(\%)$, the polarization level
of the optical continuum (UV rest frame), for all radio galaxies at $z\ga$2 for which
both have been measured (Vernet et al. \citeyear{ver01}, Dey et al. \citeyear{dey97}, Jannuzi et al. \citeyear{jan95}, De Breuck et al. in prep.). 
There is a clear trend for objects with higher Ly$\alpha$/HeII values
to show low polarization levels. In particular, except for 1243+036,
which breaks the trend, the 
LAEs (solid
circles) show the lowest polarization
levels\footnote{The polarization measurement
for 1243+036 might not be reliable, since it could be affected by strong sky residuals. 
The error bar represents a statistical estimate of the error based on noise properties
and possible defects due to sky residuals are not included.}.

The correlation  in Fig.~5 is similar to that found by 
\cite{ver01}
 on a sample of radio galaxies at 2$\la z <$3 between
Ly$\alpha$/CIV and $P(\%)$ (all the objects in their sample are
included in Fig.~5). 
Since the difference in polarization level between objects is most
likely due to the diluting effect of young stars (Vernet et
al. \citeyear{ver01}), 
Fig. 5 strongly suggests that objects with  higher
Ly$\alpha$/HeII ratios contain a relatively more luminous stellar
population. It is important to keep in mind that the
range of Ly$\alpha$/HeII ratios must also be influenced by Ly$\alpha$
absorption (it provides a natural explanation to why this ratio is 
well below standard PL model predictions in some objects).
However, since  some LAEs need an extra
supply of soft ionizing photons to enhance the Ly$\alpha$  emission, 
in spite of this uncertainty,
Fig. 5 strongly supports that stars provide such supply. 

Additional evidence for star formation
has  been found in several  LAEs. This is the case of 
 4C41.17 (Dey et al. \citeyear{dey97}, Dunlop et al. \citeyear{dun94}, Chini \& Kr\"ugel \citeyear{chi94}) and 1338-1942 (Zirm et al. \citeyear{zirm05}).
B3 0731+438,
the only LAE in our study at $z<$3 for which optical polarization measurements
exist, (Vernet et al. \citeyear{ver01}),
shows null or very low  polarization level ($<$2.4\%), suggesting the presence
of young stars.

 De Breuck et al. (in prep.) found a tentative trend 
(although more objects need to be studied to confirm this result) 
 for radio galaxies at $z\ga$3   to
show lower optical continuum polarization level than radio galaxies
in the 2$\la z <$3 range. This suggests that
 there is a relatively more luminous young
stellar population  in the highest $z$ radio galaxies.
Further   evidence comes from the dramatic increase in the
 detection rate at submm wavelengths of  $z>$2.5 radio galaxies ($\sim$75\%)  compared with $z<$2.5 radio galaxies 
($\sim$15 \%) (Archibald et al. \citeyear{arch01}).
The authors also find that the average submm luminosity rises at a rate $\propto$(1 + $z$)$^3$ out to $z\sim$4.
 Since most $z>$3 radio galaxies  are LAEs, and provided that the submm emission is  due to dust  heated by stars (e.g. Tadhunter et al. \citeyear{tad05}, Reuland et al. \citeyear{reu04})
this adds further support to our interpretation that these 
 contain
more luminous young stellar populations.

Enhanced Ly$\alpha$ ratios, well above standard PL model predictions, have
been measured
 at tens of
kpc from the nucleus for the  LAEs 4C41.17 and 1243+036  and
some radio galaxies at 2$\la z <$3 (\S4.1), most of which are not LAEs.  This
suggests that star formation is spatially extended over large spatial scales.
Spatially extended star formation in high $z$ radio galaxies has
been tentatively suggested by submm studies (Stevens et al. \citeyear{ste03}).  
For the low $z$ objects, although the contribution of the stars to the Ly$\alpha$ luminosity is not evident in the spatially integrated emission line spectrum, this becomes
apparent at tens of kpc from the nuclear region.

\begin{figure}
\includegraphics{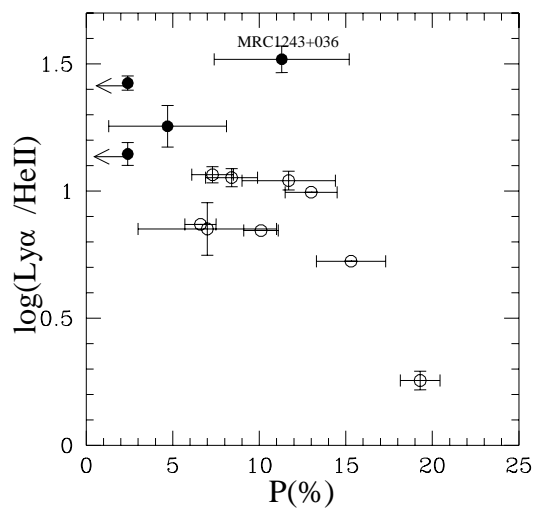}
\vspace{2.5in}
\caption{log(Ly$\alpha$/HeII) vs. percentage of polarization of the optical
continuum (UV rest frame) for all radio galaxies at $z\ga$2 for which
both have been measured. Black solid symbols correspond to LAEs, 
while hollow symbols are  objects with no Ly$\alpha$ excess.
Errorbars are shown when available. Notice the clear trend for objects with the highest Ly$\alpha$/HeII ratios to show lower polarization, as expected if the Ly$\alpha$ enhancement
is due to stellar photoionization. The trend is broken by 1243+036, which
shows too high $P(\%)$ for its large Ly$\alpha$/HeII ratio (but see text).}
\end{figure}

 Two more important issues need to be answered: Why is star
formation more intense in LAEs? Was this class of objects 
much more common at higher ($z\ge$3) redshifts and why?

The detailed spectroscopic study by \cite{hum06a} of a sample of 10 HzRG ($z=$2-3) show that  small radio sources  experience
stronger jet-gas interactions  (see also Best et al. \citeyear{best01}) and contain more luminous 
young stellar populations.
This could be the case of  LAEs. In fact, this class of objects show
in general broad emission lines and small radio sizes (see Table 3; \cite{breu00b}, \cite{breu01}, De Breuck et al. 
\citeyear{breu00b}) suggestive of jet-gas interactions. 
These are very
clear in at least two of the most extreme LAEs at $z\ge$3: 
4C41.17 (Bicknell et al. \citeyear{bick00}) and 1243+036 (van Ojik et al. \citeyear{vo96})
The  star formation might have been
induced by the interactions, although it is not clear that this process can form
enough stars to dominate the rest-frame UV emission. A more attractive possibility
is that the systems have recently
undergone a merger event which  has triggered both the star formation and the radio/AGN activities.

According to this study, we propose to use the Ly$\alpha$ ratios (specially
Ly$\alpha$/HeII) of high redshift radio galaxies and quasars to investigate
the possible signature of star formation. Although these ratios have
been measured for many radio galaxies at different redshifts, this
has been generally done for the spatially integrated spectra, where
the effect is in general less clear.
We propose that the high Ly$\alpha$ ratios relative to other emission
lines found in the extended emission line regions of some high redshift
radio-loud quasars 
(e.g.  Heckman et al. \citeyear{heck91b})
are a signature of star formation, rather than low metallicities
as often proposed. 

Star formation could also be responsible for the Ly$\alpha$ emission detected
in regions out of the reach of the quasar continuum and the radio
structures in some high redshift radio galaxies
(e.g. Villar-Mart\'\i n et al. \citeyear{vm05}, Reuland et al. \citeyear{reu03}).

We argue next that the LAE fraction shows a genuine $z$ evolution, not  due to selection 
effects. Therefore, the
radio galaxy phenomenon seems to have been more often associated with a massive starburst at $z>$3 than
at $z<$3.

\subsection{Selection effects}

Our results suggest that LAEs were more abundant at higher redshifts. 
Whether this is a genuine $z$ evolution depends on the
real fraction of LAEs at different redshifts. 

An important problem we have when comparing the $z<$3 and $z\ge$3 samples 
are the selection effects resulting from the search techniques used to
find them and the technical limitations of the spectrographs.

The parent samples were selected based on radio properties, meaning there is no a priori 
bias towards the optical properties. However, higher power radio sources are expected
to emit stronger emission lines (Willott et al. \citeyear{wil99}), so Malmquist bias on
radio power could result on an optical bias towards strong line emitters.

Is it possible that radio power is biasing our statistics? 
As De Breuck et al
(2000) explain, the effects of Malmquist bias in their sample is alleviated
for 2$< z <$4,  with the addition of several filtered surveys with lower flux density limits.
 We have compared the  radio power at 1400 MHz
(rest frame frequency, DB00a) for the $z\ge$3 and $z<$3 samples (Table 5). The median, average and standard deviation $\sigma$ are shown. 
  According to the Kolmogorov-Smirnov test, the
probability   $P$ (Table 5) that  $z\ge$3 radio galaxies have
different intrinisc radio power  than the low $z$ sample is very low (18\%).    So, we do not expect radio power
selection criteria to introduce any bias. 

Could radio size  be biasing our statistics? 
In samples to find HzRGs, the radio size has been used as an additional high redshift
filter.  As we discussed above, since
small radio sources tend to have more intense star formation, this selection criteria
could introduce some additional bias.  However, the 2$\la z<$3 and $z\ge$3 samples are not different in terms of their radio sizes
(see \S4, Table 2). On one hand this selection criteria was
applied for only $\sim$10\% of the sources in Table 1. More importantly, 0140+3253 is the only $z\ge$3 radio galaxy from such samples. 
The other 12 $z\ge$3 radio galaxies have been found without radio size constraints. This selection can therefore not explain the different LAE fractions at
 $z<$3 and z$\ge$3. 

We conclude that no bias is introduced due to radio selection criteria.

Our work is  affected by the need to detect strong emission lines
since a) this was required in the original sample to measure the redshift
b) we need at least Ly$\alpha$ and HeII to be observed to determine whether
an object shows Ly$\alpha$ excess. 
What is clear is that, due to their
Ly$\alpha$ excess compared with non-LAEs, we are not likely to have missed
low $z$ LAEs,  taking also into account that  at low $z$, CIV and HeII are in a sensitive part of the spectrographs.
There are 26 objects in De Breuck et al (DB00a DB01) at 2$\le z<$3 which could not be classified
as LAEs or non LAEs (i.e., Ly$\alpha$/HeII
not available).  An  extreme, unlikely case of all these objects being LAEs would be required
for the fractions of LAEs in the two $z$ samples to become similar (46\% vs. 54\%)
On the contrary, we would rather expect to  have missed non-LAEs in the low $z$ sample
with weak emission lines.
This would make the fraction of LAEs in the low $z$ sample even lower.

Is it possible that the sources with no CIV and HeII  are systematically biasing our statistics in the $z\ge$3 sample?  In the De Breuck et al. (2000) sample, there are 22 objects at $z\ge3$; 11 of these could not be classified as LAE's or non-LAE's. Only 6 of these had HeII outside the observed wavelength range. The remaining 5 sources had too shallow limits on the Ly$\alpha$/CIV or Ly$\alpha$/HeII ratio to classify them. Even in the extreme case that these 11 objects were non-LAEs, the fraction of LAEs at $z\ge$3 would be
7  out of 22, or 32\%, still significantly larger than in the low $z$ sample.

Is it possible that non-LAEs are systematically missing from samples of HzRGs? In the sample of ultra steep spectrum selected HzRG candidates of De Breuck et al (2001), 13 (35\%)   sources did not yield a redshift. From these, six did not show any emission, 
while the other seven had only continuum emission down to the blue end of the spectrograph, but no emission lines. The latter sources  have to be at $z<2$ (and therefore of no
interest in the present work) based on the absence of a continuum discontinuity across Ly$\alpha$. 

The former sources could be at $z\ge$2. The emission lines may be in the observed wavelength range, but too faint to be detected.
 If they were objects with Ly$\alpha$ in the observed wavelength range, it would have 
been more likely that they would  be non-LAEs, due to their intrinsically fainter Ly$\alpha$ emission.
 In any case, these sources would add at most six objects to 
our exisiting HzRG sample. In the extreme, unlikely case that these were $z\ge$3 non-LAEs, the fraction
of high $z$ LAEs would change from 54\% to 37\%, still significantly larger than in the low $z$ sample.
If these, together with the 11 $z\ge$3 non-classified objects were $z\ge$3 non-LAEs, the fraction
would be 25\% vs. less than 10\% expected in the low $z$ sample.

We conclude that although the fraction of LAEs may be incompletely determined, 
both at 2$\la z<3$ and at $z\ge3$, the large difference between both fractions is likely
to be real and not a consequence of selection effects. An extreme, unlikely  situation would be required
 to change significantly the large difference between the LAE fractions: most  non-classified 
objects at  2$\la z<3$ should be LAEs and most non-classified objects at $z\ge$3 should be non-LAEs.

In addition, we note that all $z\ge3$ LAE's are at $z>3.5$. If we compare the $z<3.5$ and $z>3.5$ sources, the fraction of LAE's are 4 out of 52 (8\%) and 7 out of 9 (78\%), respectively. It thus seems impossible to explain such large differences as due to selection effects.

\begin{table}
\begin{tabular}{ccccc} 
\hline
 $_{P(1400)/10^{35}}$  &   Median & Average & $\sigma$  & P \\   \hline
 2$\la z <$ 3    &     1.23 & 1.60 & 1.29 \\ 
$z\ge$3   &   1.20  & 1.97 & 1.56  \\  
         &  & &  & 18\%  \\   \hline
 \end{tabular}
\caption{Comparison between the rest frame radio power at 1400 MHz (in units of 10$^{35}$ erg s$^{-1}$
Hz$^{-1}$) for the 2$\la z<$3 and $z\ge$3 samples. Radio power value taken from DB00a. The median, average  and standard deviation
values  are shown. $P$
is the  probability
that the two samples are drawn from different parent populations according to the Kolmogorov-Smirnov test.  This tests implies that there is no significant difference in
intrinsic radio power between both samples.}
\end{table}

\section{Summary and conclusions}

The behaviour of the  Ly$\alpha$/HeII, Ly$\alpha$/CIV ratios and the Ly$\alpha$ line luminosity 
has been
investigated in a sample of 48  radio galaxies at 2$\la z <$3 and  13 radio galaxies at $z\ge$3
for which Ly$\alpha$/HeII and/or Ly$\alpha$/CIV measurements are available. 

 As found by \cite{breu00a}, $z\ge$3 radio galaxies tend to show larger Ly$\alpha$ ratios (2.2 and 1.5 times higher
for Ly$\alpha$/HeII and Ly$\alpha$/CIV respectively)
and higher (3 times) Ly$\alpha$ luminosities than the low $z$ sample. The Ly$\alpha$ ratios are  consistent with or above case B  predictions of standard power law
photoionization models in at least 54\% (7 out of 13) of
the $z\ge$3 objects. This is the case for only 8\% (4 out 48) of radio galaxies at
2$\leq z <$3. We refer to all these objects with unusually enhanced Ly$\alpha$
emission as Ly$\alpha$ excess objects or LAEs.  

  LAEs  show  Ly$\alpha$/HeII, Ly$\alpha$/CIV  and Ly$\alpha$ luminosity values which are  $\sim$2-3.5  times higher
 than non-LAEs.
They have radio sizes $\sim$4.6 times smaller. 36\% of LAEs show Ly$\alpha$/HeII above the standard model predictions.
Since in general no absorption has been taken into account, this fraction could be larger.

Several possibilities have been investigated to explain the Ly$\alpha$ excess 
 in LAEs: stellar photoionization,
shocks, low metallicities, high densities and absorption/resonant scattering effects.
All scenarios but the first can be rejected with confidence: 
we propose that  a population of young
stars generate the extra supply of soft ionizing photons necessary to produce the observed
Ly$\alpha$ excess in  LAEs.  Star forming rates $\sim$200 M$_{\odot}$ yr$^{-1}$
are required.  The enhanced and intense star formation activity in LAEs could be a consequence of 
a recent merger event which has triggered both the star formation and the radio/AGN activities.
This interpretation  is strongly supported by the clear trend for objects
with lower optical continuum polarization level to show larger Ly$\alpha$/HeII ratios.  It is further
supported by the tentative trend found
by other authors for $z\ge$3 radio galaxies to show lower UV-rest frame polarization levels,
or the dramatic increase on the detection rate at submm wavelengths of $z>$2.5 radio galaxies. 

The models imply that  star  formation is occuring in 45\% (5 out of 11) of LAEs.
Since absorption has not been taken into account, this is a lower limit.

The finding of enhanced Ly$\alpha$ emission  at tens of kpc from the nuclear region
in some objects suggests that
 star formation can be extended over spatial scales of tens of kpc.

We argue that, although the fraction of LAEs may be incompletely determined, 
both at 2$\la z<3$ and at $z\ge3$,  the much larger fraction of LAEs found at $z\ge$3 is  a genuine redshift evolution
and not due to selection effects. Therefore, 
  our study suggests that the radio galaxy phenomenon was
 more often associated with a massive starburst at $z>$3 than at $z<$3.  In other words,  powerful radio galaxies (and, according to the unification model, 
powerful radio quasars),  appear in more actively star forming galaxies at $z>$3 than at $z<$3.

  Enhanced Ly$\alpha$ ratios
in the extended gas of some  $z<$3 radio galaxies suggest that star formation is also present
in these objects, although at lower levels.
Star formation could also be responsible for the Ly$\alpha$ emission 
from regions out of the reach of the quasar continuum and the radio
structures detected in some high redshift radio galaxies.

We propose to use the Ly$\alpha$ ratios (specially
Ly$\alpha$/HeII) of high redshift ($z\ga$2) radio galaxies  and the extended gas in radio
loud and radio quiet quasars   to investigate
the possible signature of star formation. 

\section*{Acknowledgments}
We thank the anonymous referee for very useful comments which helped
to improve this paper substantially and Enrique P\'erez for useful comments
on the manuscript.
We thank Mark Allen for providing predictions of emission line ratios for 
 the most recent shock models, which include high velocities of up to 1000 km s$^{-1}$.
 The work of MVM has been supported by the Spanish Ministerio de Educaci\'on y Ciencia  
and the Junta de Andaluc\'\i a through the grants AYA2004-02703 and TIC-114
respectively.  AH acknowledges support in the form of a UNAM postdoctoral
fellowship.

\end{document}